%Paper: 9112065
%From: Jerome Gauntlett <jerome@rabi.uchicago.edu>
%Date: Fri, 20 Dec 91 16:11:09 CST

\magnification=\magstep1

%\hsize=160truemm
%\hoffset=10truemm
%\rightskip 10truemm
%\leftskip 10truemm

\font\list=cmcsc10

\def\real{I\negthinspace R}

\def\half{\textstyle{1\over2}}

\def\N{Nielsen}
\def\O{Olesen}
\def\NOV{Nielsen-Olesen vortex}

\def\r3{I\negthinspace R$^3$}

% HYPHENATION

\hyphenation{
mini-su-per-space
pre-factor
tei-tel-boim
es-po-si-to
haw-king
Bala-chand-ran
}

\openup 2\jot

\nopagenumbers
\hbox{ }
\rightline {FERMILAB-Pub-91/332-A}
\rightline {EFI-91-70}
\rightline{UMHEP-361}
\rightline {December 1991}
\vskip 1truecm

\centerline{\bf EUCLIDEAN BLACK HOLE VORTICES }
\vskip 1truecm

\centerline{ Fay Dowker }
\vskip 2mm

\centerline{\it NASA/ Fermilab Astrophysics Center, Fermi National
Accelerator Laboratory}
\centerline{\it P.O.Box 500, Batavia, IL 60510, U.S.A.}

\vskip 4mm

\centerline{Ruth Gregory}
\vskip 2mm

\centerline{ \it Enrico Fermi Institute, University of Chicago,
5640 S.Ellis Ave, Chicago, IL 60637, U.S.A.}

\vskip 4mm

\centerline{ Jennie Traschen }
\vskip 2mm

\centerline{ \it Department of Physics, University of Massachussetts,
 Amherst,
MA 01002, U.S.A.}

\vskip 1.5cm
\centerline{\list abstract}
\vskip 3mm

{ \leftskip 10truemm \rightskip 10truemm

\openup -1\jot

We argue the existence of solutions of the  Euclidean
Einstein equations that correspond to
a vortex sitting at the horizon of a black hole.
We find the asymptotic behaviours, at the horizon and at
infinity, of
vortex solutions for the gauge and scalar fields
in an abelian Higgs model on a Euclidean Schwarzschild
background and interpolate between them  by integrating
the equations numerically.
Calculating the backreaction shows
  that the effect of the vortex is to cut a
slice out of the Euclidean Schwarzschild geometry.
Consequences of these solutions for black hole
thermodynamics are discussed.

\openup 1\jot

}

\vfill\eject
\footline={\hss\tenrm\folio\hss}

\noindent{\bf 1. Introduction.}

\vskip 2mm

The view that the quantum aspects of black hole physics
will play an important r{\^ o}le in leading us towards a quantum theory
of gravity has been strengthened recently, not only by the discovery that
some coset conformal field theories correspond to string theory in
two-dimensional black hole geometries$^1$,
but also by the suggestion that the more familiar
four-dimensional variety can carry ``quantum hair''$^{2,3}$. This latter
development is of particular interest to relativists, since the conventional
wisdom is that powerful theorems imply that black holes are characterised
only by their mass, angular momentum and electric charge (and other charges
that are associated with a Gauss' law). Investigating these ``no-hair''
theorems, however, shows that whilst powerful, they are not
omnipotent! In particular, the existing
`no-hair theorem' for the abelian Higgs model
with the usual symmetry
breaking potential makes restrictive assumptions about the
behaviour of the fields exterior to the horizon$^{4,5}$,
restrictions that are not obviously satisfied by all
physically interesting scenarios.
It {\it has} been shown that a black hole cannot be
the source of a non-zero, static, massive vector field$^6$ but the jury is
still out on the case where a $U(1)$ gauge field acquires a mass
through the Higgs mechanism. However since the expectation is that
in this case too, black holes cannot support non-zero massive vector
fields, apparent contradictions are of great interest since they would
limit the conditions of validity of a rigorous no-hair theorem.

It has been noted by Aryal et al.$^7$ that black holes
might have hair - quite literally - since they wrote down
the metric for a black hole with a cosmic string passing
through it. They used a distributional energy
momentum source as the string,
so one could not say with
confidence that this corresponds to a physical vortex spacetime
since such a limit is not valid for line-like
defects${}^{8}$.
However, one might find this suggestive
 that a no-hair theorem
would have to be limited to the case where no topological defects
exist, thus reducing the physical relevance of such a theorem
since defects {\it will} exist if they can exist.
It was also shown by Luckock and Moss$^9$ that
black holes could carry skyrmion hair, although they
conjectured that such solutions were unstable.

More recently, it was pointed out by Bowick et
al.$^2$ that there exists a family of Schwarzschild black hole
solutions to the Einstein-axion equations labelled by a
conserved topological charge. Thus, in some sense, such black
holes could be said to be carrying axion hair. It was
then rapidly
realised that the same fractional charge that could give
rise to enhancement of proton decay catalysis by cosmic
strings$^{10}$ could potentially be carried by black holes$^3$. The
full ramifications of this type of quantum hair have been
most eloquently argued by Coleman et al.$^{11,12}$, who suggest that
this charge might have dramatic implications for black hole
thermodynamics. Remarkably, their work implies that
even if a black hole does not carry discrete charge
its temperature is
still renormalised away from the Hawking value. This means
that if we are to believe in spontaneous symmetry breaking
and the existence of strings in nature,
then we must take into account such
renormalisation effects independently of whether
discrete charge exists or not.

All of these claims rest on the existence of a family of
`vortex' solutions which are saddle points in some Euclidean
path integral. These solutions are obviously outside the
domain of standard no-hair arguments, being Euclidean, however
they are static in the sense that the
metric is static and the energy-momentum
tensor is time-independent (though not in the restricted
sense of Gibbons$^5$) and establishing existence would
set bounds on the validity of future theorems.

In this paper we will focus on the problem of existence of
solutions of the above sort. The layout of the paper is as
follows. We begin by setting
up the general problem, discussing what is meant by a
`vortex centered on a black hole'. We then show that a perturbative
analysis is justified for weakly gravitating vortices, after which we
focus on the specific example of a complex scalar (Higgs) field
with a ``Mexican hat'' potential,
coupled to a $U(1)$ gauge field.
 We find
numerically a vortex solution on a Schwarzschild background and
describe its asymptotic behaviour.
 We calculate the back-reaction on the geometry to first order in $GT$, the
energy per unit area of the vortex (in Planck units), and
also calculate the Euclidean action of this geometry.
We calculate the expectation value of the metric in a
black hole state at a certain temperature
and derive a relation between the mass and temperature without appealing
directly to the partition function.
We also calculate the expectation value of the area of the
black hole. We draw analogies with cosmic string physics, and
discuss problems with global charge.

\vskip 4mm

\noindent{\bf 2. Einstein-matter equations: general
formalism.}
\vskip 2mm

We have said we are interested in finding vortex solutions to the
abelian Higgs model in a Euclidean black hole spacetime.
First we should discuss what we mean by a Euclidean black hole
spacetime.

Recall that a Schwarzschild black hole metric has the
form
$$
ds^2 = -\left (1-{2GM\over r} \right ) dt^2 + \left (
1-{2GM\over r} \right )^{-1} dr^2 + r^2 (d\theta^2 + \sin^2
\theta d\phi^2).
\eqno (2.1)
$$
We may formally Euclideanise this by setting $t\to
i\tau$. However, we now see that the former Lorentzian
coordinate singularity at $r=2GM$ is in danger of
becoming a real singularity in the Euclidean space,
since the metric changes signature from four to zero
for $r<2GM$. This tells us that we must regard $r>2GM$
as the only region of relevance in our Euclidean
section, and that therefore we must be able to include
$r=2GM$ in a non-singular fashion into our manifold.
Changing variables to $\rho^2 = 16G^2M^2 r^{-1}(r-2GM)$ we see that
$$
ds^2 = \rho^2 d({\tau\over 4GM})^2 + d\rho^2 + 4G^2M^2
d\Omega^2_{_{II}}
\eqno (2.2)
$$
near $r=2GM$, which shows that $\tau$ must be
identified with period $8\pi GM$, and that $r$ and
$\tau$ are analogous to cylindrical polar coordinates
on a plane. Thus, we arrive at the conclusion that
Euclidean Schwarzschild has topology
$S^2\times$\real$^2$, with a periodic time coordinate,
period $\beta=8\pi GM$. The geometry of the $t-r$ section
of Euclidean Schwarzschild can be visualised as the surface
of a semi-infinite ``cigar'' with a smoothly capped end and
tending to a cylinder of radius $4GM$ as $r\rightarrow\infty$.

In general there will be matter present as well as a
black hole, therefore, assuming that the matter is
spherically symmetric and `static' (i.e., cylindrically
symmetric), we will be looking for solutions to the
Euclidean Einstein equations with topology
$S^2\times$\real$^2$, being spherically symmetric on the $S^2$
sections, and cylindrically symmetric on the \real$^2$ sections.
(Note that we require only the energy momentum to have these
symmetries. It is quite possible that the constituent
fields do not, for example, a \NOV~is cylindrically
symmetric even though the Higgs field has a dependence on
the azimuthal coordinate.)
The metric is then a function of just one variable, a radial
coordinate in the \real$^2$ plane.
The presence of a black hole is indicated by the existence of  a
minimal value of the radial coordinate $r_s$ ($= 2GM$, say) at which
the metric and curvature are nonetheless regular. Following
Garfinkle et al.$^{13}$ we will write the metric in the form
$$
ds^2 = A^2 d\tau^2 + A^{-2} dr^2 + C^2 (d\theta^2 +
\sin^2 \theta d\phi^2)
\eqno (2.3)
$$
where
$A(r_s)=0$, $\tau$ is understood to be a periodic coordinate with
period $\beta$, and $C(r_s)^2 = {\cal A}/4\pi$ is given in terms
of the area of the event horizon. The regularity of the metric at
$r_s$ implies we can choose local cylindrical coordinates in
which the metric is regular
$$
\rho = BA(r)
\eqno (2.4)
$$
where $B=\beta/2\pi$ is used for convenience. Regularity then
implies $(A^2)'|_{r_s} = 2/B $. In principle we can leave the
metric in terms of the period, $\beta$, and the area of the event
horizon, ${\cal A}$, however, for calculational simplicity we
choose to use up the coordinate freedom
$$
r \to ar+b \;\;\; , \;\;\; \tau \to a^{-1}\tau
\eqno (2.5)
$$
to set $B=2r_s$ and $C(r_s) = r_s$. We may then re-interpret our
coordinates if required. The Einstein equations for this metric
can then be written as:
$$
\eqalignno{
C'' &= 4\pi G {C\over A^2} (T^0_0 - T^r_r) & (2.6a) \cr
((A^2)'C^2)' &= 8\pi GC^2( 2T^\theta_\theta + T^r_r - T^0_0)& (2.6b)
\cr
{2AA'C'\over C} - {1\over C^2} (1-A^2C'^2)
&= 8 \pi G T^r_r & (2.6c) \cr
}
$$
where
$$
T_{ab} = {2\over \sqrt{g} } {\partial ({\cal L} \sqrt{g})
\over \partial g^{ab}}
\eqno (2.7)
$$
is the energy momentum tensor, which obeys the conservation law
$$
{T^r_r}' + {A'\over A} (T^r_r - T^0_0) + {2C'\over C} ( T^r_r -
T^\theta_\theta) = 0
\eqno (2.8)
$$
which is valid for a general spherical-cylindrical
symmetric source.

In order to complete our preliminaries on
formulating the Einstein equations, we note that since we
expect the greatest variation of $T^a_b$ to occur near the
horizon, it may be expedient to have a form of the Einstein
equations in terms of the proper distance from the horizon.
For convenience we also scale out the dimensional fall-off
behaviour of the energy momentum tensor, $r_H$ say, to
express quantities in terms of the dimensionless parameter
$$
{\hat r} = {1\over r_H} \int_{r_s}^r {dr'\over A} \; .
\eqno (2.9)
$$
Setting ${\hat C}(\hat r)=C/r_s$, and $\epsilon {\hat T}^a_b
= 8\pi G T^a_b r_H^2$, the boundary conditions at
the horizon become
$$
{\hat C}(0) = 1, \;\;\; {\hat C}'(0) = 0, \;\;\;
{\hat C}''(0) = {1\over 2R^2}
+ {\half} \epsilon {\hat T}^0_0 |_{r_s}\; ,
\eqno (2.10a)
$$
and
$$
A(0) = A''(0) = 0 \; , \;\;\; A'(0) = {1\over 2R}\; ,
\eqno (2.10b)
$$
where prime denotes ${{d}\over{d \hat r}}$ and $R = r_s/r_H$
is the ratio of the Schwarzschild radius to the vortex
width. The Einstein equations are now
$$
\eqalignno{
(A'{\hat C}^2)' &= \epsilon {\hat C}^2A
( 2{\hat T}^\theta_\theta + {\hat T}^r_r -
{\hat T}^0_0)& (2.11a) \cr
\left ( {{\hat C}'\over A} \right ) '
&= {\half} \epsilon {{\hat C}\over A} ({\hat T}^0_0
- {\hat T}^r_r) & (2.11b) \cr
{\hat C}' &= {-A'{\hat C}\over A} \left [ 1 - \sqrt{ 1 + {A^2 \over A'^2
{\hat C}^2} \left ( {1\over R^2} + \epsilon {\hat C}^2 {\hat T}^r_r \right )
} \right ] \; , & (2.11c) \cr
}
$$
where we have rearranged (2.6c) as a quadratic for ${\hat C}'$.
Regularity at the horizon fixes the sign of the root in
(2.11c), which is then valid in some neighbourhood of the horizon.

Having set up this formalism, we now turn to the problem of
deciding under what circumstances we expect a vortex black
hole to exist.

\vskip 4mm

\noindent{\bf 3. Asymptotic solution of Einstein's equations. }

\vskip 2mm

We would like to show that solutions exist which correspond
to a vortex at the horizon of the black hole. However,
rather than taking a specific field theory source for
$T^a_b$, in this section we remain more general,
investigating what minimal conditions $T^a_b$ must satisfy in order
to have an asymptotically Schwarzschild metric. We
naturally have in mind that $T^a_b$ has some, as yet
unspecified, field theory vortex solution as its source,
therefore we expect $T^a_b=E{\hat T}^a_b/r_H^2$,
where $E$ is an energy per unit area
characterising the source, ${\hat T}^a_b$ is the rescaled
energy momentum referred to in (2.11) which is of order
unity, and $r_H$ represents a
cut-off scale of the vortex. Thus, for example, a
\NOV~has $E \sim \eta^2$ and $r_H \sim
1/\sqrt{\lambda}\eta$, where $\eta$ is the symmetry
breaking scale and $\lambda$ the quartic self-coupling
constant. Because we are in Euclidean space, we do not have
a conventional set of energy conditions for $T^a_b$, but
since we know that $T^a_b$ is derived from a $\theta$ and
$\phi$ independent field theoretic
lagrangian, we do have a modified dominant energy condition,
namely that
$$
{\cal L} = - T^\theta_\theta = - T^\phi_\phi \geq |T^0_0|,
|T^r_r|.
\eqno (3.1)
$$

Now, as we have already remarked, we are looking for a non-singular
asymptotically Schwarzschild metric. This means that we do
not expect $C=0$, nor in fact do we expect $A'=0$ at any
finite $r$. (We cannot make a similar statement concerning
$C'$, since the effect of the radial stresses can conspire
to make $C$ actually {\it decrease} near the horizon.)
Inspection of (2.11a) shows that
$A'({\hat r})>0$ is guaranteed if
$$
J({\hat r}) = \epsilon \int_0^{\hat r} {\hat C}^2 A (2{\hat
T}^\theta_\theta + {\hat T}^r_r - {\hat T}^0_0)d{\hat r}'
\eqno (3.2)
$$
converges, and its modulus is less than $1/2R$.
What we will now prove
is that if $\epsilon=8\pi GE \ll 1$ (the vortex is suitably
weakly gravitating) and if the energy momentum satisfies
certain fall-off conditions then $J$ is not only convergent,
but is of order $\epsilon/R$. By a fall-off condition we
mean that outside the core (${\hat r} \geq $few)
$|{\hat T}^a_b| \leq K ({\hat r}^{-n})$ for some $K$ of
order unity, $n>0$. Our aim is to find a value of $n$ which
will guarantee that we can integrate out the metric
functions to large values of ${\hat r}$. This will then tell
us what sort of energy momenta we expect well-behaved vortex
solutions to have. Since we are not, at this stage, trying
to argue the existence of a full solution to the coupled
Einstein-matter system, we restrict our attention to only
two of the metric equations, (2.11a,c). The reason for this
is that the three Einstein equations implicitly contain the
matter equations of motion, conservation of energy momentum
being an integrability condition for (2.11a-c). Now let us
turn to proving our claim - and finding the value of $n$.

We start by assuming the contrary - that $J$ is divergent.
Then there exists an ${\hat r}_0$ at which $J({\hat r}_0) =
-1/4R$, thus on $[0, {\hat r}_0]$ (2.11a) implies
$$
{1\over 2R} \geq A' {\hat C}^2 \geq {1\over 4R}.
\eqno (3.3)
$$
Now, in order to use (2.11c) to bound ${\hat C}$, we must be
sure that the sign of the root is fixed; this relies
crucially on
$$
f({\hat r}) = {A'^2 {\hat C}^2 \over A^2} + {1\over R^2} +
\epsilon {\hat C}^2 {\hat T}^r_r
\eqno (3.4)
$$
being positive. Let ${\hat r}_f \leq {\hat r}_0$ be chosen
so that $f>0$ on $[0, {\hat r}_f]$. Then, on this interval
$$
-\sqrt{\epsilon} {\hat C} |{\hat T}^r_r|^{{\half}}
\leq {\hat C}' \leq \left ( {1\over R^2} + \epsilon {\hat
C}^2 |{\hat T}^r_r| \right ) ^{1\over 2}
\eqno (3.5)
$$
using $\{ 1 - \sqrt{|y|} \leq \sqrt{1 + x + y} \leq 1 +
\sqrt{ x + |y|} \}$ for $x>0$, $|y|<1$.

Let us consider the implications of each bound in turn. The
lower bound on ${\hat C}'$ implies
$$
{\hat C} \geq \exp \{ - \sqrt{\epsilon} \int |{\hat
T}^r_r|^{1\over 2} \} \geq e^{-\alpha\sqrt{\epsilon}}
\eqno (3.6)
$$
where $\alpha$ will be order unity if we use the fall-off
assumption with $n\geq 4$, (and so in particular $\hat C$ is always
positive). Hence
$$
A' \leq {1\over 2R} e^{2\alpha \sqrt{\epsilon}} \;\;\;\;\;
\Rightarrow \;\;\;\;\;
A \leq {{\hat r}\over 2R} e^{2\alpha \sqrt{\epsilon}} \;\;\;\;\;
{\rm on} \; [0, {\hat r}_f].
\eqno (3.7)
$$
Using this bound and (3.3) we see that
$$
{A'^2 {\hat C}^4 \over A^2} +
\epsilon {\hat C}^4 {\hat T}^r_r \geq {e^{-2\alpha
\sqrt{\epsilon}}\over 4{\hat r}^2} - \epsilon
e^{-4\alpha\sqrt{\epsilon}} |{\hat T}^r_r|
\eqno (3.8)
$$
is strictly positive on $[0,{\hat r}_f]$ provided $\epsilon
\ll 1$ and the previous fall-off assumption holds. Therefore
${\hat C}^2 f > {\hat C}^2 /R^2$ on $[0,{\hat r}_f]$,
and without loss of generality, we may choose ${\hat
r}_f = {\hat r}_0$.

Now we examine the upper bound on ${\hat C}$:
$$
{\hat C}' \leq {1\over R} + \sqrt{\epsilon} {\hat C} |{\hat
   T}^r_r|^{1\over 2} \leq
{e^{\sqrt{\epsilon}\int |{\hat T}^r_r|^{1\over 2}} \over R}
+  \sqrt{\epsilon} {\hat C} |{\hat T}^r_r|^{1\over 2} ,
\eqno (3.9)
$$
which implies that
$$
{\hat C} \leq e^{\sqrt{\epsilon}\int |{\hat T}^r_r|^{1\over 2} }
( 1 + {{\hat r} \over R} ).
\eqno (3.10)
$$
Bounding $\int |{\hat T}^r_r|^{1\over 2} $ by $\alpha$ as
before, we see that
$$
|J| \leq \epsilon \int_0^{\hat r} {{\hat r}\over 2R}
e^{4\sqrt{\epsilon}\alpha} ( 1 + {{\hat r}\over R} )^2
|2{\hat T}^\theta_\theta + {\hat T}^r_r - {\hat T}^0_0| d{\hat r}.
\eqno (3.11)
$$
This is readily seen to be convergent on
$[0, {\hat r}_0]$ if $n\geq 5$ in the
fall-off assumption, and we may write
$$
|J| \leq {\epsilon\gamma \over 2R}
\eqno (3.12)
$$
for some $\gamma$ of order unity provided $R\geq 1$. Therefore
for $R\geq 1$, $J({\hat r}_0)$ cannot be equal to $1/4R$
thus contradicting the initial assumption about ${\hat
r}_0$. Therefore we conclude that no such ${\hat r}_0$
exists, and provided that $|{\hat T}^a_b| \leq K {\hat
r}^{-5}$ we may (formally) integrate out the metric
equations to infinity keeping $A', {\hat C} >0$.
Note again that this argument only involves (2.11a)
and (2.11c).

We now use the following argument to conclude
that if a solution does exist then it is asymptotically
Schwarzschild.

Note that the initial conditions imply that
$\int_{r_s}^{r_s + \delta} (r-r_s)|T^0_0 - T^r_r|/A^2dr$ is
bounded. But then we use $A > A(r_s+\delta)$
on $(r_s+\delta,\infty)$ to conclude that
$$
\int_{r_s+\delta}^\infty {(r-r_s)|T^0_0 - T^r_r| \over A^2}dr
< {E \over 4RA(r_s+\delta)} \int_{{\hat r}(r_s+\delta)}^\infty
{\hat r}^2 |{\hat T}^0_0-{\hat T}^r_r| d{\hat r} < \infty.
\eqno (3.13)
$$
We may then use a theorem\footnote{$^\dagger$}{The
theorem states that if $\int_0^\infty x|a(x)|dx$ is
bounded, then the non-zero solutions of the
 2$^{\rm nd}$ order equation $u'' +
a(x)u = 0$ have the asymptotic form $u\sim Ax+B$
where the constants $A$ and $B$ cannot both be zero$^{14}$.}
from ordinary differential equations to
conclude that
$$
C \sim cr + d \;\;\; {\rm as}\; r\to\infty.
\eqno (3.14)
$$
Examining (2.6b,c) as $r\to \infty$ shows that $c\neq 0$ and
(2.6b) then implies $(A^2)' \to 0$ as
$r \to \infty$, and a
rearrangement of (2.6c) gives
$$
A^2 \sim {1\over c^2} \left ( 1 - {(r_s+I)\over r}\right )
\;\;\; {\rm as}\;\; r\to \infty,
\eqno (3.15)
$$
where
$$
I = 8\pi G \int_{r_s}^r C^2 (2T^\theta_\theta + T^r_r
- T^0_0) dr' = r_s R J.
\eqno (3.16)
$$

Thus we see that any solution must be asymptotically Schwarzschild.
We can also see that the solution will be changed by $O(\epsilon)$
from exact Schwarzschild. Indeed,
$$
2AA'C^2 = r_s + I ( = r_s ( 1 + O(\epsilon) ))
\eqno (3.17)
$$
implies
$$
{C\over A^2} ( T^0_0 - T^r_r ) = 2 { (C^3 T^r_r)' - C^2C'
(T^r_r + 2T^\theta_\theta) \over (r_s + I) }
\eqno (3.18)
$$
using the equations of motion for $T^a_b$. Then, using (2.6c) at
the horizon to determine $C'|_{r_s} = 1 + 8\pi G r_s^2
T^r_r|_{r_s}$, we may rewrite (2.6a) as
$$
\eqalignno{
C'(r) &= 1 + {\epsilon C^3 T^r_r \over (r_s +I) E}
+ {\epsilon \over E} \int {C^2 (T^r_r + 2T^\theta_\theta) \over
(r_s + I)} \left [ {\epsilon C^3 T^r_r \over (r_s+I) E} - C'
\right ] dr - {\epsilon^2 \over E^2} \int {C^5 T^r_r T_0^0
\over (r_s+I)^2} dr \cr
& \to 1 - {\epsilon \over Er_s} \int_{r_s}^\infty C^2 (T^r_r +
2T^\theta_\theta) dr + O(\epsilon^2) \;\;\; {\rm as} \; r \to
\infty & (3.19) \cr
}
$$
which gives the value of $c$ to order $\epsilon$.

It is possible to write integral expressions for
the changes in the Arnowitt-Deser-Misner (ADM)
mass$^{15}$ and period of the spacetime from their
vacuum values.
Recall from (3.14,15) that the asymptotic form of the metric
is
$$
ds^2 = c^{-2} \left ( 1 - {r_s + I \over r} \right ) d\tau^2
+ c^2 \left ( 1 - {r_s +I \over r} \right ) ^{-1} dr^2 +
(cr+d)^2 d\Omega^2_{_{II}}.
\eqno (3.20)
$$
where $c$ is given by (3.19).
If $c\neq 1$, then clearly the $\tau,r$ coordinates
are not those of a `Euclidean observer' at infinity. In
order to identify the true period and ADM mass
of the space, we must rescale the $r,\tau$
coordinates so that $A^2\to1$ at infinity. Thus we set
$$
\tau' = \tau /c \;\;\; ; \;\;\; r' = cr +d
\eqno (3.21)
$$
to obtain
$$
ds^2 = \left ( 1 - {(r_s + I)c \over r'-d} \right ) d\tau'^2
+ \left ( 1 - {(r_s +I)c \over r'-d} \right ) ^{-1} dr'^2 +
r'^2 d\Omega^2_{_{II}},
\eqno (3.22)
$$
and hence
$$
\beta' = \beta /c
= \beta \left ( 1 + {\epsilon \over Er_s}
\int_{r_s}^\infty C^2 (T^r_r +
2T^\theta_\theta) dr \right )
\eqno (3.23)
$$
$$
M_\infty = c(r_s+I(\infty))/2G
= {r_s \over 2G} \left ( 1 - {\epsilon \over r_s E}
\int_{r_s}^\infty C^2 T^0_0 dr \right ) \; .
\eqno (3.24)
$$
are the period and ADM mass of the space to order $\epsilon$.

Thus to order $\epsilon$, the period of the geometry
decreases, whereas $M_\infty$ may increase or decrease
according to the details of the specific vortex model
chosen.

The preceeding expressions give the modified period
and ADM mass of
the spacetime, if one knows what the solutions are.
However, a perturbation expansion in $\epsilon$ for solutions
is justified if $\epsilon \ll 1$  and we will now give
the solutions for the metric functions in the perturbative case.
One can solve for the sources
$T^a _b (r)$ as test fields on the Schwarzschild background. In the next
section we will study the equations for the matter fields in the Abelian
Higgs model, so for now let us assume that we have solved the equations
and know what the vortex
 sources are. These solutions on the background are exact if
$\epsilon=0$, i.e. the matter and gravity
decouple.
The next step is to compute the corrections
to the metric coefficients when $\epsilon \neq 0$.

One finds that to first order
$$
C=C_1=r + \int_{r_s}^r dr' I_1(r')
\eqno(3.25)
$$
where
$$
I_1 (r)=
{\epsilon\over {E r_s}} \left [ r^3 T^r_r -
\int _{r_s}^r dr' r'^2(T^r_r+2T^\theta_\theta) \right ] ,
\eqno(3.26)
$$
and
$$
A^2 =A_1^2=1 - {{r_s}\over r} +\int_{r_s}^r dr'
\left({{I(r')}\over{{r'}^2}} -{{2r_s}\over{{r'}^3}}
\int_{r_s}^{r'} ds I_1(s)\right)
\eqno(3.27)
$$
where $I(r)$ is given by (3.16) with $C$ replaced by
$r^2$.
In equations (3.25)
and (3.27)
 everything on the right hand side is known, in terms of
the sources.

For large $r$ one can then extract the derivative of $C$ and the ADM
mass, to give the modifications to the period and mass
which are just equations (3.23) and (3.24) with the metric
functions in  the integrals replaced by their Schwarzschild
forms.

\vskip 4mm

\noindent{\bf 4. An abelian Higgs vortex solution.}

\vskip 2mm

We now  examine the specific energy momentum source of an
abelian Higgs
vortex centered on the horizon.
 The lagrangian for the matter fields is
$$
{\cal L} =  \left ( {1\over 4} F_{\mu\nu}^2 +
({\cal D}^\mu \psi )^* {\cal D}_\mu \psi +
%({\cal D}^\mu \phi )^* {\cal D}_\mu \phi +
{\lambda \over 4}
(|\psi|^2 - \eta^2)^2 \right ) \; .
\eqno (4.1)
$$
For a simple vortex solution we choose
the variation of the phase of the $\psi$
field to distribute itself uniformly over the periodic $\tau$
direction. This is simply a gauge choice which
allows us to simplify the
equations of motion by setting
$$
\eqalign{
\psi &= \eta X(r) e^{ik\tau/B} \cr
A_\mu &= {1\over Be} ( P(r) - k)\partial_\mu \tau = {1\over Be}
(P_\mu - k\partial_\mu \tau) \; . \cr
}
\eqno (4.2)
$$
This implies
that the lagrangian and equations of motion simplify to
$$
{\cal L} = \left \{ {P_{,r}^2 \over 2e^2B^2}
+ \eta^2 X_{,r}^2 A^2 + \eta^2 {X^2P^2\over A^2B^2} +
{\lambda \eta^4 \over 4} (X^2-1)^2 \right \}
\eqno (4.3)
$$
$$
\eqalignno{
{1\over C^2} ( C^2 P_{,r})_{,r} &= {\lambda \eta^2 \over \nu} {X^2 P
\over A^2} & (4.4a) \cr
{1\over C^2} (C^2 A^2 X_{,r})_{,r} &= {P^2 X \over A^2 B^2} +
{\lambda \eta^2\over 2} X(X^2 -1)\; , & (4.4b)   \cr
}
$$
where $\nu = \lambda /2e^2$.

It is
straightforward to check that the asymptotic behaviour of
the bounded solutions to (4.4) is
$$
X \propto (r-r_s)^{|k|/2} \hbox{\hskip 5mm} P = k - \alpha(r-r_s)
\hbox{\hskip 5mm \rm as} \; r\to r_s \eqno (4.5a)
$$
where $\alpha=-{Be/(4\pi r_s^2)}\int_H A'_\tau dS$
and
$$
1-X \propto r^{-1} e^{-\sqrt{\lambda}\eta r/A_\infty}
\hbox{\hskip 5mm} P
\propto r^{-1} e^{-\sqrt{\lambda} \eta r/\sqrt{\nu} A_\infty}
\hbox{\hskip 5mm \rm as}
\; r\to\infty  \eqno(4.5b)
$$
where $A_\infty$ is given by (3.15).
The appearance of the square root in the dependency of $X$ on $r$
near the horizon simply reflects the dependence on the local
proper distance there. Note that at this level, there is no
obvious obstruction to the fall-off condition on ${\hat
T}^a_b$ being satisfied.

If solutions to the coupled Einstein-Higgs equations
exist, then we expect that there is a perturbative limit as
$\epsilon\to 0$, as we have
noted\footnote{$^\ddagger$}{This limit might seem problematic
since it involves taking either $G\to 0$ or $E=\eta^2\to 0$.
The former limit must be taken at finite $r_s$ in order to
preserve the background geometry, this would mean that
the Euclidean black hole would have a formally infinite
``mass''. The latter limit is
equivalent to sending the symmetry breaking scale
to zero which would require sending the self coupling,
$\lambda$ and the charge, $e$, to infinity in order to keep
$r_H$ and $\nu$ fixed. Since, by rescaling the fields,
 one can express the equations in terms of
$\epsilon$, $r_H$ $r_s$ and $\nu$ only, both limits are
equivalent as far as the equations are concerned. However,
since G is a measured physical constant, it may be
easier to think of the limit as $\eta \to 0$.}. Indeed,
many of the demonstrations of the lack of `hair' on
Lorentzian black holes have shown that on a fixed
Schwarzschild background the interaction between test fields
and a source is extinguished as the source approaches the
horizon$^{16}$. Therefore we first consider the question of the
existence of solutions for the matter fields on a fixed
Euclidean Schwarzschild background, setting $C=r$ and $A^2 =
1-{r_s\over r}$ in (4.4). Rescaling the radial variable to
${\tilde r} = {r-r_s\over r_H}$ gives
$$
\eqalign{
{1\over ({\tilde r} +R)^2}
\left [ ({\tilde r} +R)^2 P' \right ] ' &=
{X^2P\over \nu} {( {\tilde r} +R) \over {\tilde r}} \cr
{1\over ({\tilde r} +R)^2}
\left [ {\tilde r} ({\tilde r} +R) X' \right ] ' &=
{P^2X\over 4R^2} {( {\tilde r} +R) \over {\tilde r}}
+ {\half} X (X^2-1) \cr
}
\eqno (4.6)
$$

The question is - is there a solution to (4.6) which connects the
bounded behaviour at the horizon (4.5a) to the bounded behaviour at
infinity (4.5b with $A_{\infty}=1$)?
Existence of such solutions is similar
to the difficult question of existence of abelian Higgs
vortices in Minkowski spacetime, first investigated by
Nielsen and Olesen${}^{17}$. To see this, consider flat space
and make the
ansatz (4.2) with $\rho$ and $\theta$ replacing $r$
and $\tau$ respectively, where $\rho$ and $\theta$ are
cylindrical polar coordinates in the plane perpendicular to
the infinitely long straight static \NOV. Setting $X_{\rm
NO}$ and $P_{\rm NO}$ as the \N-\O~solutions, the equations
of motion that these satisfy can be readily seen to be
$$
\eqalign{
({\rho} X_{NO}')' &= {X_{NO}P_{NO}^2 \over {\rho}} + {\half}{\rho}
X_{NO}(X_{NO}^2-1) \cr
\left ( { P_{NO}'\over{\rho} } \right ) ' &=
{X_{NO}^2P_{NO} \over \nu{\rho} }
\cr
}
\eqno (4.7)
$$
Existence of solutions to these equations was shown
numerically by Nielsen and Olesen, and their stability
properties discovered by Bogomoln'yi$^{18}$. Much is known about the
behaviour of Nielsen-Olesen vortices, or cosmic strings. In particular,
Bogomoln'yi showed that for a special value of $\nu$,
$\nu=1$, the second order equations in (4.7) reduce to two
first order equations:
$$
\rho X_{NO}' = X_{NO}P_{NO}
 \;\;\;\;\; ; \;\;\;\;\;\; P_{NO}'/\rho = {\half} X_{NO}
(X_{NO}^2-1).
$$
This is often referred to as the supersymmetric limit, since
the model is supersymmetrisable for this value of $\nu$.
The above relations also have the direct consequence that
the radial and azimuthal stresses, $T^\rho_\rho$,
$T^\theta_\theta$, vanish identically. For $\nu\neq1$, these
stresses become non-zero changing sign according to the
value of $\nu$. This idea will be important in our later
discussions of the mass and entropy. However, for the
moment, let us just note that for $\nu \leq 1$ vortex
solutions are stable for all values of the winding number
$k$, whereas for $\nu>1$, solutions with $k\geq2$ are
unstable.

In order to see the similarities (and differences) between
our problem and the \N-\O~case we have just discussed, let
$z=\rho^2/4R$, then (4.7) becomes
$$
\eqalign{
{1\over R} [ z X,_z ],_z &= {XP^2 \over 4Rz} + {\half} X
(X^2-1) \cr
P,_{zz} &= {R\over z} {X^2P \over \nu} \cr
}
\eqno (4.8)
$$
The two sets of equations (4.6) and (4.8) become identical
as $z,{\tilde r} \ll R$.
However, far from the horizon,
$z,{\tilde r} \gg R$, the equations are very different, and we
cannot simply infer the existence of well-behaved solutions
to (4.6) from the \N-\O~case.

We do not currently have an analytic proof of the existence
of regular solutions to (4.6), however, we have integrated
the equations numerically using a relaxation technique, and
these results show that the bounded eigenfunctions at the
horizon do indeed integrate out to the exponentially decaying
eigenfunctions at infinity. Figure 1 shows a plot of $X$ and
$P$ with $k=1$, $\nu=1$ and $R=2$, compared with the Nielsen-Olesen
solutions. The radial coordinate is ${\tilde r}$ for the
Schwarzschild case and $\rho$ for the \N-\O~case.
The difference in behaviours at the origin reflects the fact that
for the Schwarzschild case $r$ is not the coordinate in which the
metric near the horizon looks flat.
At $r=0$, $X'_{SCHW}=\infty$, $P'_{SCHW}={-1.92}$,
$X'_{NO}={1.37}$ and $P'_{NO}=0$.

Having justified the existence of a background solution, let
us remark on the behaviour of a fully coupled system.
Setting
$$
{\hat \rho} = \rho/r_H = 2RA(r),
$$
a local cylindrical coordinate, we find
$$
\eqalignno{
{1\over {\hat\rho}} { r_s^2 (r_s+I)^2 \over C^4}
({\hat\rho} X')' &= {XP^2 \over {\hat\rho}^2 } + {\half} X(X^2-1) -
\epsilon {\hat\rho} X' (2 {\hat T}^\theta_\theta + {\hat T}^r_r -
{\hat T}^0_0 ) & (4.9a) \cr
{1\over {\hat\rho}} {r_s^2 (r_s+I)^2 \over C^4}
(P'/{\hat\rho})' &= {X^2P \over \nu{\hat\rho}^2 } -
\epsilon {P'\over {\hat\rho}} (2 {\hat T}^\theta_\theta + {\hat T}^r_r -
{\hat T}^0_0 ) , & (4.9b) \cr
}
$$
or, alternatively
$$
{\hat\rho} ({\hat T}^r_r)' + ( {\hat T}^r_r - {\hat T}^0_0 )
+ [ O(\epsilon) + O(\hat \rho^2 R^{-2}) ] ( {\hat T}^r_r - {\hat
T}^\theta_\theta ) =0,
\eqno (4.10)
$$
where $I=O(r_s\epsilon)$ is given by (3.16).

Now, noting that $C = r_s ( 1 + O(\epsilon) + O(R^{-2}) )$
for ${\hat\rho} \ll R $,
from (3.6,10), we readily see the similarity of (4.9) with
(4.7). We also see that the matter equations can be written as some
background piece plus an order $\epsilon$ piece coming from
the interaction of the vortex with the geometry.
This then justifies the iterative procedure for the matter
part of the fully coupled system.

To zeroth order, the space is Euclidean Schwarzschild,
$$
C=r,\;\; A^2 = \left ( 1- {r_s\over r} \right ) , \;\; {\hat
\rho} = 2R \sqrt{ 1 - {r_s\over r} } \; .
\eqno (4.11)
$$
In order to calculate the back reaction we will focus on thin
vortices, since these are more physically relevant.
This limit corresponds to $R\gg1$, and we therefore expect
our solutions to be well approximated by the Nielsen-Olesen
solution for ${\hat \rho} \ll R$, of the exponential
form (4.5b) for ${\hat\rho} > R$, and having some
transitionary nature from ${\hat\rho}$-exponential decay to
$r$-exponential decay for intermediate radii. We will in
fact assume $R^{-2} \ll \epsilon$ to facilitate the following
analysis, keeping in mind that for a typical GUT vortex
$\epsilon \sim 10^{-6}$
would only require $r_s \gg 10^3 r_H \sim
10^{-26}$cm! Since $R$ is so very large, the energy momenta
are negligibly small for ${\hat \rho} \geq R$, so as far as
the Einstein equations are concerned we can essentially
ignore corrections from the Nielsen-Olesen form for
${\hat\rho} \geq R$ as well, and we will simply set
$$
X_0 = X_{\rm NO} (\hat\rho) \;\;\; ; \;\;\;
P_0 = P_{\rm NO} (\hat\rho)
\eqno (4.12)
$$
where it is understood that $X_0$ and $P_0$ have
$O(R^{-2})$ corrections which do not
contribute to the order in perturbation theory ($O(\epsilon)$)
to which we will be working.

The results of section 3 allow us to now calculate the back-reaction on
the metric quite straightforwardly. In what follows we will
suppress the suffix $0$ on the energy momentum
tensor for clarity. Setting
$$
{\hat\mu} = -\int {\hat\rho}
{\hat T}^\theta_\theta d{\hat\rho}
\eqno (4.13)
$$
the normalised energy per unit area of the vortex, and
${\hat p} =
-\int {\hat\rho} {\hat T}^r_r d{\hat\rho}$, an averaged
scaled pressure,
we see that (3.18) implies that, to first order in $\epsilon$,
$$
C'(\infty) = 1 + \epsilon ({\hat\mu} + {\half} {\hat p}).
\eqno (4.14)
$$
Then, noting from (4.10) that
$$
\int {\hat\rho} ({\hat T}^0_0 + {\hat T}^r_r) d{\hat\rho} = O(\epsilon),
\eqno (4.15)
$$
the ADM mass parameter from (3.23) is
$$
M_\infty = {r_s\over 2G} ( 1 - {\half}\epsilon {\hat
p} )
\eqno (4.16)
$$
to first order in $\epsilon$.
Thus, making the coordinate transformation defined in
(3.21):
$$
r' = (1-\epsilon({\hat\mu} + {\half} {\hat p})) r \;\;\;\; ;
\;\;\; \tau' = (1-\epsilon({\hat\mu} + {\half} {\hat p})) \tau
\eqno (4.17)
$$
the asymptotic metric takes the form
$$
ds^2 = \left ( 1 - {2GM_\infty \over r'} \right )
d\tau'^2 +
\left (  1 - {2GM_\infty \over r'} \right )
^{-1} dr'^2 + r'^2 d\Omega_{_{II}}^2 \; .
\eqno (4.18)
$$
Therefore our asymptotic solution takes the form of
Schwarzschild, with an adjusted period
$$
\eqalign{
\beta' &= \beta ( 1 - \epsilon ({\hat\mu} + {\half} {\hat p}) )\cr
&= 8\pi G M_\infty ( 1 - \epsilon {\hat\mu}) \cr
}
\eqno (4.19)
$$
and mass parameter $M_\infty$, adjusted that is, relative to
the `expected' mass-period relationship derived at the
horizon. Note also that the area of the black hole is now
related to the ADM mass via
$$
{\cal A} = 4\pi r_s^2 = 16\pi G^2 M_\infty^2 (1 + \epsilon
{\hat p}).
\eqno (4.20)
$$

Note some similarities with a self-gravitating cosmic
string. There the \real$^2$
sections perpendicular to the string acquire an asymptotic
`deficit angle'$^{18}$ $\delta\theta = -(2\pi).4 G\mu$, where
$$
\mu = 2\pi\eta^2 \int {\hat \rho} {\hat T}^\rho_\rho
d{\hat\rho} = 2 \pi\eta^2 {\hat\mu}
\eqno (4.21)
$$
is the energy per unit length of the cosmic string in its
rest frame.
Here we see that our `deficit angle' is $\delta \tau = -8\pi
G M_\infty \epsilon {\hat\mu} = - (8\pi GM_\infty).4G\mu$.
Since we expect the period of $\tau$ to be $8\pi GM_\infty$
(as we expect the $2\pi$ period in $\theta$), we see that
the form of the correction in both cases is the same. Thus,
the gravitational effect of the vortex is to `cut' a wedge
or slice out of the Euclidean black hole cigar outside the
vortex. In figure 2 we show a schematic representation of
the black hole vortex geometry.

As we remarked at the end of the previous section, the vortex
always decreases the period compared to its Schwarzschild value
for a black hole of a given horizon area. The ADM mass, on
the other hand, can be larger than, smaller than or equal
to its Schwarzschild
value for fixed horizon area, depending on $\hat p$.
Existing results for
 a self gravitating cosmic string${}^{19}$
indicate that for $\nu > (<)\; 1$, $\hat p > (<)\; 0$. These
results were numerically obtained and so may only be true
to a certain order, however they indicate that there is some
critical value of $\nu$, close to 1, for which
the average pressure, $\hat p$,
changes sign. Now, in our case, the background is flat space
only to zeroth order in $R^{-2}$ so we expect that the critical
value of $\nu$, $\nu_{C}$, differs from the flat space
value by $O(R^{-2})$ and thus is still close to 1.

\vskip 4mm

\noindent{\bf 5. Actions, temperature and entropy.}

\vskip 2mm

Having calculated the gravitational effect of the vortex, it
is instructive to calculate the  Euclidean action:
$$
I_E = \int \left \{ {\cal L}_{ M} - {{\cal R}\over
16\pi G} \right \} \sqrt{g} d^4x - {1\over 8\pi G} \int_{\Sigma}
(K-K^0) \sqrt{h} d^3x
\eqno (5.1)
$$
where $K$ is the trace of the extrinsic curvature of
$\Sigma$ - a boundary ``at infinity'', calculated in the
true geometry and $K^0$ the extrinsic curvature trace
calculated for $\Sigma$ isometrically embedded in flat space. For our
asymptotically flat geometry, $C \sim r'$, $A^2 = 1 -
{2GM_\infty \over r'} + O(r'^{-2})$, this boundary term has
the value
$$
I_\Sigma = {\half} \beta' M_\infty.
\eqno (5.2)
$$
For the pure vortex source, we may use the Einstein equations
to deduce
that the Ricci scalar ${\cal R} = 16\pi G {\cal L}_{ M} - 8\pi G
(T^r_r + T^0_0)$. However, from (4.15) we see that
$$
\int C^2 (T^r_r + T^0_0) dr ={1\over G} O(\epsilon^2).
\eqno (5.3)
$$
Thus
$$
\int \left \{ {\cal L}_{M} - {{\cal R}\over
16\pi G} \right \} \sqrt{g} d^4x = {1\over G} O(\epsilon^2) .
\eqno (5.4)
$$
Therefore, we come to the conclusion
that, to first order in $\epsilon$,
 the Euclidean action is, as with Schwarzschild, equal
to its boundary term, ${1\over 2} \beta' M_\infty$. However,
reading off the relation between $\beta'$ and $M_\infty$
from (4.19), we see that
$$
I_E = {\beta'^2 \over 16\pi G} ( 1 + \epsilon {\hat\mu}) +
{1\over G}O(\epsilon^2)
\eqno (5.5)
$$
in terms of the period. However, note
that
$$
{\beta'^2 \epsilon {\hat\mu} \over 16\pi G}
= {-\beta'^2 \over r_s} \int_{r_s}^\infty C^2
T^\theta_\theta dr = {\beta' \over 4\pi r_s } \int
{\cal L}_{M} \sqrt{g} d^4x= \int
{\cal L}_{M} \sqrt{g} d^4x + O(\epsilon^2)\; .
\eqno (5.6)
$$
Hence
$$
I_E(\beta') = {\beta'^2 \over 16\pi G} + \int {\cal L}_{M}
\sqrt{g} d^4 x = I_0(\beta') + I_M(\beta'),
\eqno (5.7)
$$
to first order in $\epsilon$,
where $I_0(\beta')$ is the action of Schwarzschild with period $\beta'$
and $I_M(\beta')$ is the action of the $X_0$, $P_0$ solution in
the background of Schwarzschild with period $\beta'$.
Therefore, taking into account the back-reaction of the
vortex on the geometry, we confirm the value of the
Euclidean action used by Coleman et al.$^{11}$

The interest of computing the Euclidean vortex solutions is that
their actions contribute to the gravitational path integral. In the
path integral one must decide which fields to include in the sum.
One prescription is to include all metrics and
matter fields with a particular
fixed period, $\beta$, and this describes "a system at temperature
$1/\beta $". Here we compute what follows from such a prescription.
Other boundary conditions are possible, which will be explored in
further work.

Having calculated the vortex geometry we are in a
position to directly calculate the expectation value of
the mass of a black hole of temperature $1/\beta$ using
$$
<g_{ab}> =(1 + \sum_{\pm}C_{\pm}e^{-I_{\pm}})^{-1}[ g_{_0ab} +
\sum_{\pm}C_{\pm} e^{-I_\pm} g_{_\pm ab}] + O(
e^{-2I_\pm})
\eqno (5.8)
$$
where $g_{_0ab}$ is the Schwarzschild metric with period $\beta$,
$g_{_+ ab }=g_{_- ab}$ are the $k=\pm 1$ vortex geometries with period
$\beta$ and $I_{+}=I_{-}$ are the matter parts of their actions.
$C_{+}=C_{-}$ are the determinants of quadratic fluctuations about
the vortices.

This formula is derived from a Euclidean path integral and must be used
with caution since the metric is not a gauge invariant quantity.
One must add the metrics at the same
point of the space-time manifold,
which concept has no diffeomorphism
 invariant meaning. However,
in this case, since the metrics are all asymptotically flat, we can fix
coordinates in the asymptotic region and only use the formula
(5.8) there. In each case we choose coordinates such that
$g_{00}\to 1$, and the area of the two-spheres is $4\pi r^2$
as a function of $r$ at infinity.

Since the geometries for $k=\pm 1$ are identical, setting
$C=C_+ + C_-$ yields
$$
\eqalign{
<g_{00}> &\sim (1 + Ce^{-I_M})^{-1}
 \left ( 1 + Ce^{-I_M} - {{2G}\over{r}}(M + Ce^{-I_M}
M_\infty)\right ) \cr
<g_{rr}>  &\sim (1 + Ce^{-I_M})^{-1}
 \left ( 1 + Ce^{-I_M} + {{2G}\over{r}}(M + Ce^{-I_M}
M_\infty)\right ) \cr
<g_{\theta\theta}> &={{<g_{\phi\phi}>}\over{\sin^2 \theta}}
\sim r^2 \cr}
\eqno (5.9)
$$
as $r\to \infty$, where $I_M=I_{\pm}$ and
$$M={\beta\over {8\pi G}},\;\;\;M_\infty={\beta\over{8\pi G}}(1+
\epsilon \hat\mu).$$

Substituting in for the masses we obtain
$$\eqalign{
<g_{00}>\sim 1 -  {\beta\over {4\pi r}}(1+ \epsilon \hat\mu Ce^{-I_M})
\cr
<g_{rr}>\sim 1 +  {\beta\over {4\pi r}}(1+ \epsilon \hat\mu Ce^{-I_M}).
\cr}
\eqno (5.10)
$$
Thus we have
$$<M(\beta)>={\beta\over{8\pi G}}[1+Ce^{-I_M} \epsilon\hat\mu]
\eqno(5.11)
$$
as the predicted value of the mass of a black hole with temperature
$\beta^{-1}$.
Noting that, for $k=\pm 1$,  $\epsilon {\hat\mu} = 4 T_{\rm string}$ in the
notation of Coleman et al., this is readily
seen to agree with their expression for the modified Hawking
temperature of the black hole$^{11}$.

The horizon is another place where we can make sense of (5.8).
It is a two-sphere and for each metric in (5.8) we know its
area, $\cal A$, in terms of the period, giving
$$
<{\cal A}> = {{\beta}^2\over 4\pi} \left [ 1 +
Ce^{-I_M} (2\epsilon{\hat\mu} + \epsilon{\hat p} ) \right ]
\eqno (5.12)
$$
for the expectation value of the area of the black hole.
We compare this with the entropy, $S(\beta)$,
calculated from the partition function, $Z(\beta)$, via
$$S=\beta^{2} {\partial\over{\partial \beta}}\left(-\beta^{-1}
{\rm{ln}}Z\right).\eqno (5.13)
$$
Approximating the Euclidean path integral for $Z(\beta)$
semiclassically yields
$$Z(\beta)=e^{{-\beta^2}\over{16\pi G}}\left(1+Ce^{-I_M}\right)
\eqno (5.14)$$
and thus
$$
4GS(\beta) =
{{\tilde\beta}^2\over 4\pi} \left [ 1 +
2\epsilon{\hat\mu} Ce^{-I_M} \right ] - Ce^{-I_M}.
\eqno (5.15)
$$
We find that the central formula $S = {1\over 4G}A$ in black
hole thermodynamics has now apparently been violated, and depending on
the specifics of the vortex (i.e. the size and sign of $\hat p$) S
can either be greater than or less than ${1\over 4G}<A>$.
Note that the result (5.12) could not be obtained
from the partition function since it contains an $\epsilon {\hat p}$ term.

\vskip 4mm
\noindent{\bf 6. Conclusions. }

\vskip 2mm

To summarise: we have argued the existence of solutions of the
coupled Einstein-vortex equations by showing that under suitable
fall-off conditions of the energy-momentum of a
weakly gravitating vortex a perturbative analysis
is justified. We have demonstrated a suitable vortex for
beginning an iterative procedure by numerically obtaining
a vortex solution of the abelian Higgs model in a Schwarzschild
background.
 We calculated the mass-period-area relations for the corrected
geometry to first order in $\epsilon$, the gravitational
strength of the vortex and used these results
to derive the renormalised mass of a black hole of a certain
temperature.
We also found that the expected value of the horizon area
is not related to the entropy of the black hole in the
usual way.

Our work also provides a potential `no-go' argument for
global vortices. In the cosmic string
scenario, local strings have asymptotically conical spacetimes whereas
static global string spacetimes are singular$^{21}$, the energy
momentum tensor having only a $1/r^2$
fall-off in flat space. In our Euclidean case, the energy
of a global vortex in the Schwarzschild background
would have no fall-off due
to the fixed circumference ($\beta$) of
$r,\theta,\phi=$const circles. Therefore, drawing an analogy between
these two situations, if static global cosmic
strings are singular we do not expect
global black hole vortices to be otherwise. Not having asymptotically
flat geometries, they would therefore not contribute to
the partition function.

We mentioned the effect of varying the parameter $\nu$ on the
results obtained. For the flat space Nielsen-Olesen
vortex, the critical value of $\nu$ is exactly 1.
In that case, $\nu>1$ means that
a string with winding number $k\ge 2$ is unstable${}^{17}$,
alternatively, that the vortices repel one another, whereas
$\nu<1$ implies that they attract. Since we have argued that
just such a critical value of $\nu$, $\nu_{C}$ close to 1, exists for
the black hole vortices, it is interesting
to speculate that, for
$\nu>\nu_{C}$, the $k\ge 2$ solutions are unstable,
i.e. are not minima of the Euclidean action.
In that case the $k\ge 2$ solutions that we have found would
not contribute to a Euclidean path integral.
It seems plausible to suppose that stable solutions of
the matter equations on a Schwarzschild background do
exist, which would consist of two separate string world sheets
sitting opposite each other ($\tau_2-\tau_1 = {1\over 2}\beta$)
at finite distance from the horizon, where any further loss of energy
due to moving further away would be balanced by an increase
in energy due to increase in the area of the world sheets.
Such a solution would not be cylindrically
symmetric and its action would differ from the form calculated in
(5.6), although presumably the difference would be small.
However, it would be interesting to investigate
such types of solutions.

Our derivation of the geometry not only enabled us to
confirm the results of Coleman et al., but we were also able to
calculate the expected area of the black
hole. We obtained what looks
to be a discrepancy in the usual area-entropy
relationship, though, in this case, virtual
string world sheets ``dress'' the black hole around the horizon and
perhaps one should not expect the area-entropy relation to survive.
However, it is
the pressure, rather than some combination of energy and
pressure, that is contributing to the
discrepancy and this result certainly
merits further thought.

\vskip 4mm

\noindent{\bf Acknowledgements. }

\vskip 2mm

We would like to thank Peter Bowcock, Gary Gibbons,
Stephen Hawking  and Robert Wald for useful
discussions. We are especially grateful to Jerome
Gauntlett and David Kastor for assistance during the
early stage of this work and to Jeff Harvey for useful
suggestions.
F.D.~was supported in part by the DoE and NASA grant
NAGW-2381 at Fermilab, R.G.~by the NSF grant PHY 89-18388
and the McCormick Fellowship fund at the Enrico Fermi
Institute, University of
Chicago and J.T.~by the NSF.

\vskip 4mm

\noindent{\bf References.}

\vskip 2mm

\parskip=0pt

% MACROS

\newcount\refno
\refno=0
\def\nref#1\par{\advance\refno by1\item{[\the\refno]~}#1}

\def\totalno{\message{***Total:\the\refno ***}}

\def\book#1[[#2]]{{\it#1\/} (#2).}

\def\annph#1 #2 #3.{{\it Ann.\ Phys.\ .(N.\thinspace Y.) .\bf#1} #2 (#3).}
\def\apj#1 #2 #3.{{\it Ap.\ J.\ \bf#1} #2 (#3).}
\def\cmp#1 #2 #3.{{\it Commun.\ Math.\ Phys.\ \bf#1} #2 (#3).}
\def\cqg#1 #2 #3.{{\it Class.\ Quantum Grav.\ \bf#1} #2 (#3).}
\def\foundph#1 #2 #3.{{\it Found.\ Phys.\ \bf#1} #2 (#3).}
\def\grg#1 #2 #3.{{\it Gen.\ Rel.\ Grav.\ \bf#1} #2 (#3).}
\def\intjmp#1 #2 #3.{{\it Int.\ J.\ Mod.\ Phys.\ \rm A\bf#1} #2 (#3).}
\def\jmp#1 #2 #3.{{\it J.\ Math.\ Phys.\ \bf#1} #2 (#3).}
\def\jphysa#1 #2 #3.{{\it J.\ Phys.\ \rm A\bf#1} #2 (#3).}
\def\mpla#1 #2 #3.{{\it Mod.\ Phys.\ Lett.\ \rm A\bf#1} #2 (#3).}
\def\mnras#1 #2 #3.{{\it Mon.\ Not.\ R.\ Ast.\ Soc.\ \bf#1} #2 (#3).}
\def\nat#1 #2 #3.{{\it Nature\ \bf#1} #2 (#3).}
\def\ncim#1 #2 #3.{{\it Nuovo Cim.\ \bf#1\/} #2 (#3).}
\def\ncimb#1 #2 #3.{{\it Nuovo Cim.\ \bf#1\/}B #2 (#3).}
\def\np#1 #2 #3.{{\it Nucl.\ Phys.\ \bf#1} #2 (#3).}
\def\npb#1 #2 #3.{{\it Nucl.\ Phys.\ \rm B\bf#1} #2 (#3).}
\def\phrep#1 #2 #3.{{\it Phys.\ Rep.\ \bf#1} #2 (#3).}
\def\pl#1 #2 #3.{{\it Phys.\ Lett.\ \bf#1} #2 (#3).}
\def\plb#1 #2 #3.{{\it Phys.\ Lett.\ \bf#1\/}B #2 (#3).}
\def\pr#1 #2 #3.{{\it Phys.\ Rev.\ \bf#1} #2 (#3).}
\def\prb#1 #2 #3.{{\it Phys.\ Rev.\ \rm B\bf#1} #2 (#3).}
\def\prd#1 #2 #3.{{\it Phys.\ Rev.\ \rm D\bf#1} #2 (#3).}
\def\prl#1 #2 #3.{{\it Phys.\ Rev.\ Lett.\ \bf#1} #2 (#3).}
\def\pcps#1 #2 #3.{{\it Proc.\ Cambs.\ Phil.\ Soc.\ \bf#1} #2 (#3).}
\def\plms#1 #2 #3.{{\it Proc.\ Lond.\ Math.\ Soc.\ \bf#1} #2 (#3).}
\def\prs#1 #2 #3.{{\it Proc.\ Roy.\ Soc.\ \rm A\bf#1} #2 (#3).}
\def\revmod#1 #2 #3.{{\it Rev.\ Mod.\ Phys.\ \bf#1} #2 (#3).}
\def\rprog#1 #2 #3.{{\it Rep.\ Prog.\ Phys.\ \bf#1} #2 (#3).}
\def\sovpu#1 #2 #3.{{\it Sov.\ Phys.\ Usp.\ \bf#1} #2 (#3).}
\def\sovjpn#1 #2 #3.{{\it Sov.\ J.\ Part.\ Nucl.\ \bf#1} #2 (#3).}
\def\sovj#1 #2 #3.{{\it Sov.\ J.\ Nucl.\ Phys.\ \bf#1} #2 (#3).}
\def\jtpl#1 #2 #3.{{\it Sov.\ Phys.\ JETP\ Lett.\ \bf#1} #2 (#3).}
\def\jtp#1 #2 #3.{{\it Sov.\ Phys.\ JETP\ \bf#1} #2 (#3).}

\nref
E.Witten, \prd 44 314 1991.

\nref
M.Bowick, S.Giddings, J.Harvey, G.Horowitz and A.Strominger, \prl 61 2823 1988.

\nref
L.Krauss and F.Wilczek, \prl 62 1221 1989.

\nref
S.Adler and R.Pearson, \prd 18 2798 1978.

\nref
G.Gibbons, \book Self gravitating magnetic
monopoles, global monopoles and black holes [[R 90/31]]

\nref
J.Bekenstein, \prd 5 1239 1972.

\nref
M.Aryal, L.Ford and A.Vilenkin, \prd 34 2263 1986.

\nref
R.Geroch and J.Traschen \prd 36 1017 1987.

\nref
H.Luckock and I.Moss, \plb 176 341 1986.

\nref
Ph.de Sousa Gerbert, \prd 40 1346 1989.

M.Alford, J.March-Russell and F.Wilczek, \npb 328 140 1989.

\nref
S.Coleman, J.Preskill and F.Wilczek, \mpla 6 1631 1991.

\nref
S.Coleman, J.Preskill and F.Wilczek, \prl 67 1975 1991 1991.

\nref
D.Garfinkle,G.Horowitz and A.Strominger, \prd 43 3140 1991.

\nref
R.Bellman, \book Stability Theory of Differential Equations
[[McGraw-Hill, New York, 1953]]

\nref
R.Arnowitt, S.Deser and C.Misner in \book Gravitation: an Introduction
to Current Research [[ed. L.Witten, Wiley, New York, 1962]]

\nref
C.Teitelboim, \prd 5 2941 1972.

J.B.Hartle, \prd 1 394 1970.

\nref
H.B.Nielsen and P.Olesen, \npb 61 45 1973.

\nref
E.B. Bogomoln'yi, \sovj  24 449 1976.

\nref
D.Garfinkle, \prd 32 1323 1985.

R.Gregory, \prl 59 740 1987.

\nref
P.Laguna-Castillo and R.Matzner, \prd 36 3663 1987.

\nref
R.Gregory, \plb 215 663 1988.

G.Gibbons, M.Ortiz and F.Ruiz-Ruiz, \prd 39 1546 1989.

\bye